\documentstyle[epsfig]{aipproc}

\newcommand{\apj}{ApJ}

\newcommand{\mnras}{MNRAS}
\newcommand{\pasj}{PASJ}
\newcommand{\pasp}{PASP}

\begin{document}

\title{\textit{Beppo}SAX Observations of Mkn~421: \\
clues on the particle
acceleration ? }

\author{G. Fossati$^1$, A. Celotti$^2$, M. Chiaberge$^2$ and Y.H. Zhang$^2$}
\address{$^1$ UCSD/CASS, 9500 Gilman Drive, La Jolla, CA 92093-0424, U.S.A. --- gfossati@ucsd.edu \\
$^2$ SISSA, via Beirut 2--4, 34014 Trieste, Italy --- celotti, chiab, yhzhang@sissa.it}

\maketitle

\begin{abstract}
Mkn~421 was repeatedly observed with \textit{Beppo}SAX in 1997--1998.
We present highlights of the results of the thorough temporal
and spectral analysis discussed by Fossati et al. (1999) and Maraschi et
al.  (1999), focusing on the flare of April 1998, which was simultaneously
observed also at TeV energies.
The detailed study of the flare in different energy bands
reveals a few very important new results:
(a) hard photons lag the soft ones by 2--3 ks --a behavior opposite to
what is normally found in High energy peak BL Lacs X--ray spectra;
(b) the flux decay of the flare can be intrinsically achromatic if 
a stationary underlying emission component is present.
Moreover the spectral evolution during the flare has been followed 
by extracting X--ray spectra on few ks intervals, allowing to 
detect for the first time the peak of the synchrotron component shifting to
higher energies during the rising phase, and then receding. 
The spectral analysis confirms the delay in the flare at the higher
energies, as above a few keV the spectrum changes only after the peak of
the outburst has occurred.
The spectral and temporal information obtained challenge the simplest
models currently adopted for the (synchrotron) emission and most
importantly provide clues on the particle acceleration process.  
A theoretical picture accounting for all the observational constraints
is discussed, where electrons are injected at low energies and
then progressively accelerated during the development of the flare.

\end{abstract}


\section*{Introduction}
\label{sec:introduction}

Blazars are radio--loud AGNs characterized by strong variability,
large and variable polarization, and high luminosity. 
The spectral energy distribution (SED) typically shows two broad peaks in a
$\nu F_\nu$ representation (Fossati et al. 1998), with the emission up to
X--rays thought to be due to synchrotron radiation from high energy
electrons, while it is likely that $\gamma$-rays derive from the same
electrons via inverse Compton (IC) scattering.
In X--ray bright BL Lacs (HBL, from High-energy-peak-BL Lacs, Padovani
\& Giommi 1995) the synchrotron maximum occurs in the
soft-X--ray band, and the IC emission extends in some
cases to the TeV band.

\underbar{Mkn~421} ($z$ = 0.031) is the brightest HBL at X--ray
and UV wavelengths and the first extragalactic source discovered at TeV
energies (Punch et al. 1992), where dramatic variability has been observed
(Gaidos et al. 1996).

\section*{The 1998 X--ray/TeV flare}

\begin{figure}[b]
\centerline{\includegraphics[width=0.55\linewidth]%
{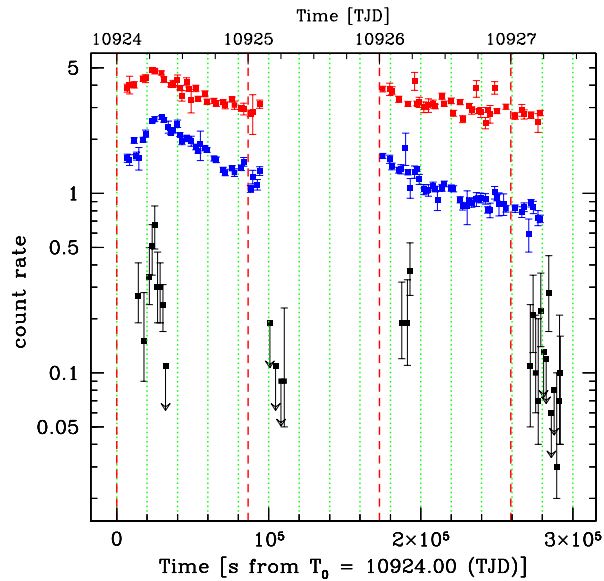}}
\caption{
Light curves of Mkn~421 at TeV and X--ray energies, during the 1998 campaign.  
They are shown in order of increasing energy from bottom to top: Whipple
E$\geq$ 2~TeV, MECS 4.0--6.0~keV and LECS 0.1--0.5~keV (both with 1500 s
bins, multiplied by a factor 4 and 8, respectively).  The count rate
units are cts/s for \textit{Beppo}SAX data, and cts/min for Whipple
data.  
\label{fig:lc98tev} }
\end{figure}

In 1998 \textit{Beppo}SAX observed Mkn~421 as part of a long lasting
monitoring campaign (see also Takahashi in these Proceedings).
\textit{Beppo}SAX observations are dominated by an isolated flare (see
Fig.~\ref{fig:lc98tev}), and one of the striking and important results is
that in correspondence with the X--ray flare of April 21$^{\rm st}$ a sharp
TeV flare was detected by the Whipple Cherenkov Telescope
(Figure~\ref{fig:lc98tev}).  The peaks in the 0.1--0.5~keV, 4.0--6.0~keV
and 2~TeV light curves are \textit{simultaneous within one hour}
(see Maraschi et al. 1999).

Here we will focus on the X--ray characteristics of the April 21$^{\rm st}$
flare.  
We accumulated light curves for different energy bands.
The post--flare light curves have been modeled with an exponential decay,
superimposed to a steady emission.
Four the main results:  

\settowidth{\labelwidth}{A}
\hangindent \labelwidth
\hangafter 1
\noindent
\underbar{Decay Timescales:}
the timescales range between 30 and 45$\times 10^3$ seconds, and \textit{do
not} show a clear (if any) relationship with the energy, rather suggesting
that the post--flare spectral evolution can be \textit{achromatic}.
This result leads to reject the simplest possibility that the decay
evolution is driven by the radiative cooling of emitting electrons (this
simplest picture would produce a dependence of the timescale with energy,
$\tau \sim {\rm E}^{-1/2}$).

\hangindent \labelwidth
\hangafter 1
\noindent
\underbar{Flaring/Steady components:}
exponential decay fits require the presence of an underlying less variable
component.

\hangindent \labelwidth
\hangafter 1
\noindent
\underbar{Time Lag:}
the harder X--ray photons lag the soft X--ray ones.
We performed a cross correlation analysis using the DCF (Edelson \& Krolik
1988) and the MMD (Hufnagel \& Bregman 1992) techniques 
and statistically determined the significance of the time lags using Monte
Carlo simulations (Peterson et al. 1998).
We refer to Zhang et al. (1999) for the relevant details of such analysis.
The result is an average lag of $-2.7^{+1.9}_{-1.2}$ ks for DCF,
and $-2.3^{+1.2}_{-0.7}$ ks for MMD (1 $\sigma$).
This finding is opposite to what is normally found in the best
monitored HBL X--ray spectra (e.g. Urry et al. 1993; Kohmura et al. 1994; 
Takahashi et al. 1996; Zhang et al. 1999) whose hard-to-soft behavior is
usually interpreted in terms of cooling of the synchrotron emitting particles. 

\hangindent \labelwidth
\hangafter 1
\noindent
\underbar{Rise vs. Fall:}
possible ``asymmetry'' of the rise/decay of the flare
especially for the higher energy X--rays.  The flare seems to be symmetric
at the energies corresponding (roughly) to the synchrotron peak, while it
might have a faster rise at higher energies.
This could be connected to the observed hard--lag.

\section*{Spectral Variability (1997 \& 1998)}

We accumulated spectra in sub-intervals, and developed an
\textit{intrinsically curved spectral model} to be able to estimate the
position of the peak of the synchrotron component.
In 1997 the source was in a lower brightness state, with a softer 
($\Delta\alpha_{97,98} \simeq 0.4$) X--ray spectrum at all
energies, and the peak energy 0.5~keV lower.
There is a clear relation between the flux variability and the changes in
the spectral parameters, both in 1997 and in 1998.

\begin{figure}[t]
\centerline{%
\includegraphics[width=0.48\linewidth]{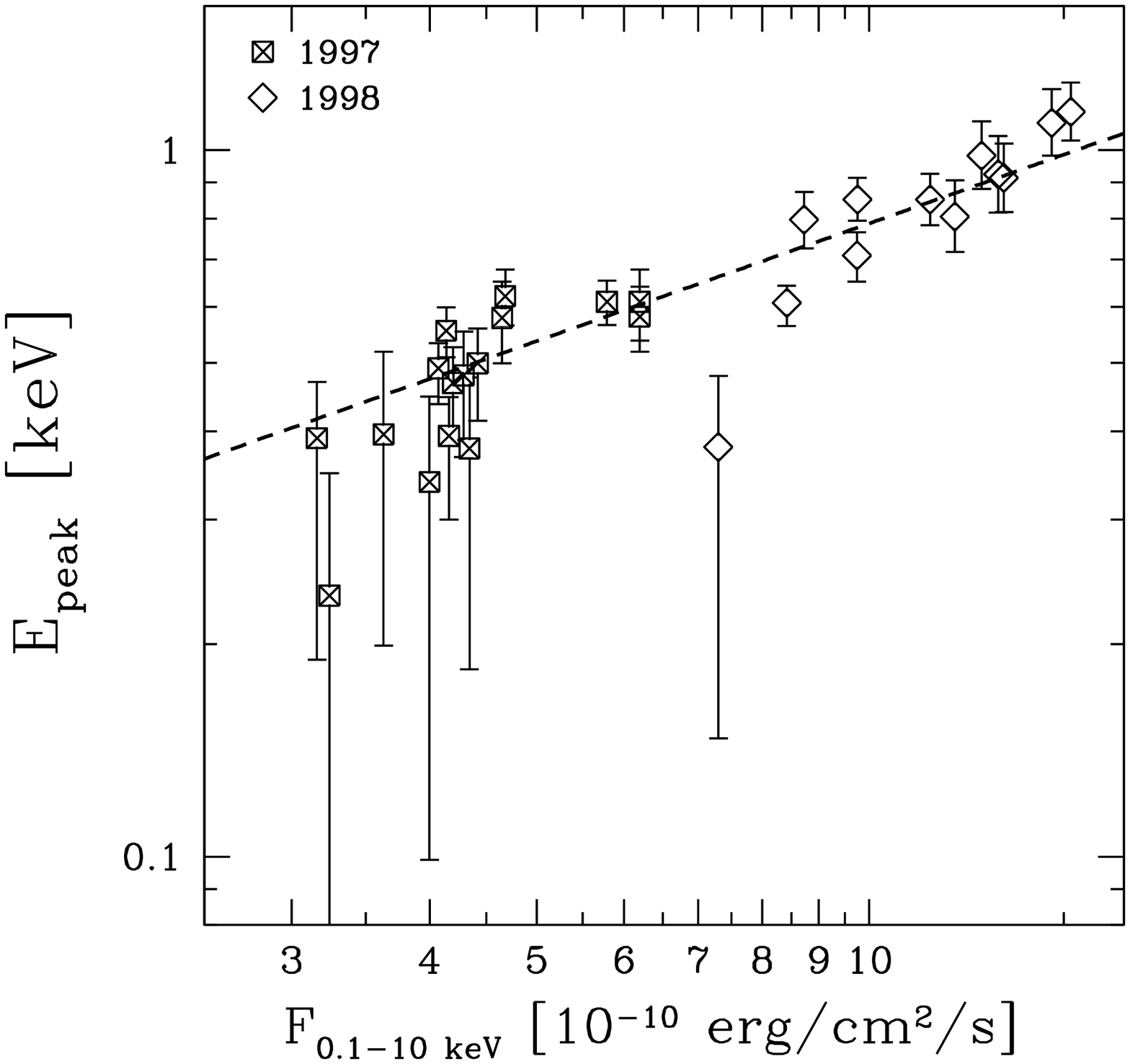}\hfill%
\includegraphics[width=0.49\linewidth]{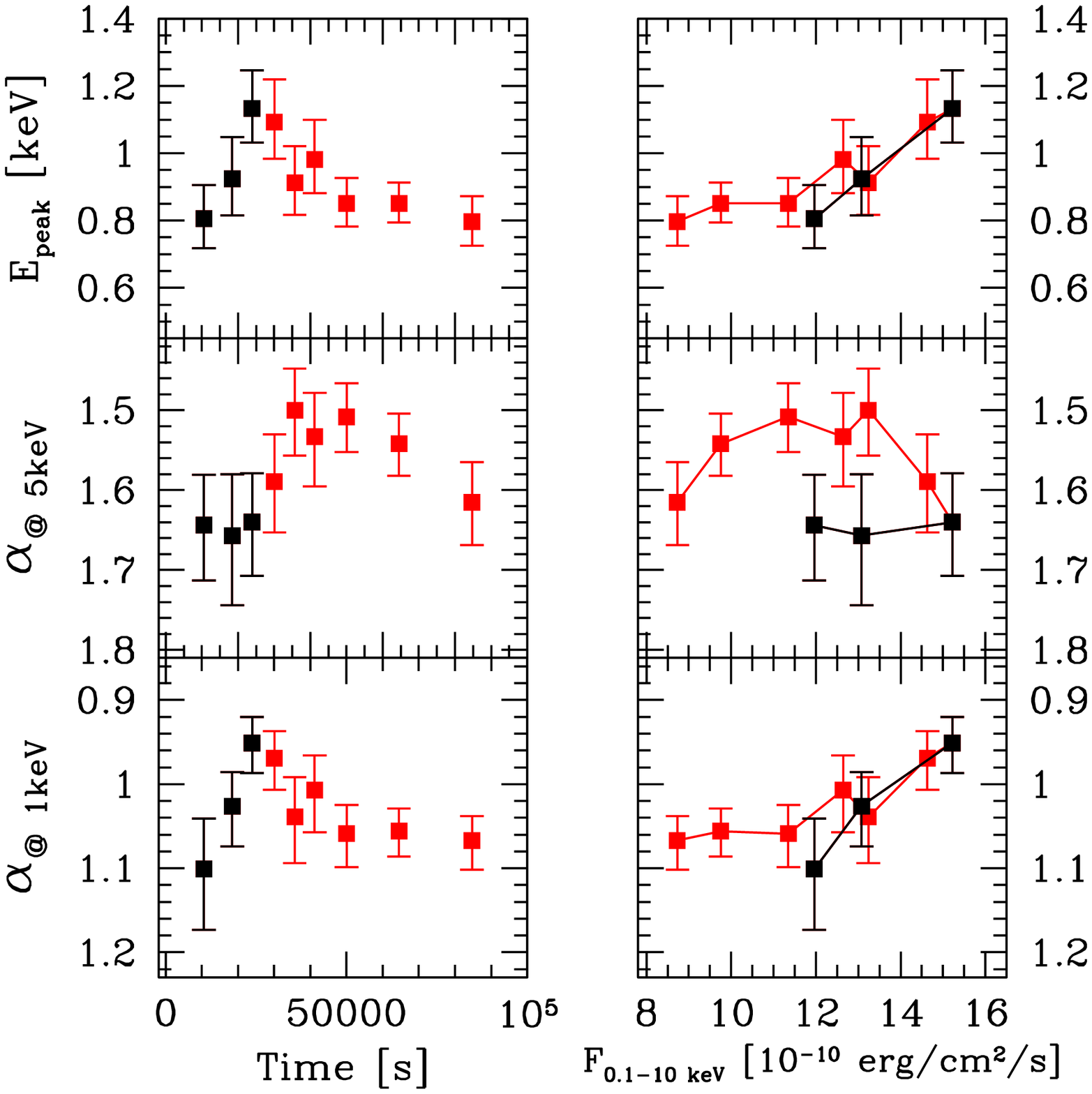}
}
\vspace{-0.5cm}
\begin{minipage}[t]{0.49\linewidth}
\caption{
Synchrotron peak energy plotted versus ``de-absorbed'' 0.1--10.0~keV flux.
The dashed line represents the best fitting power
law, having a slope $\epsilon=0.55$.
\label{fig:epeak_vs_Flux} }
\end{minipage}
\hfill
\begin{minipage}[t]{0.49\linewidth}
\caption{
The photon spectral indices at 1~keV and at 5~keV, and energy of the peak
of the synchrotron component, are plotted versus time and ``de-absorbed''
0.1--10.0~keV flux. 
\label{fig:dettaglio_98} }
\end{minipage}
\end{figure}

\hangindent \labelwidth
\hangafter 1
\noindent
\underbar{Synchrotron peak energy:}
the main new result is that we were \textit{able to determine the
energy of the peak of the synchrotron component} (with its error).
We find a correlation between changes in the brightness and shifts of
the peak position (e.g. Fig.~\ref{fig:epeak_vs_Flux}).  
The source reveals a strikingly coherent spectral
behavior between 1997 and 1998, and through a large flux variability (a
factor 5 in the 0.1--10.0~keV band).
The peak energies lie along a tight relation E$_{\rm peak} \propto {\rm
F}^{0.55 \pm 0.05}$.

\hangindent \labelwidth
\hangafter 1
\noindent
\underbar{Hard Lag in 1998 spectra:}
the spectral analysis confirms the signature of the hard lag.
A blow up of the 1998 flare interval is shown in
Figure~\ref{fig:dettaglio_98}.  
The main remarkable features are:
(a) the synchrotron peak shifts toward higher energy during the rise, and
then decreases as soon as the flare is over.  
(b) The spectral index at 1~keV reflects exactly the same behavior, as
expected being computed at the energy around which the peak is moving.  
(c) On the contrary, the spectral shape at 5~keV does not vary until a
few ks after the peak of the flare, and only then --while the flux is
decaying and the peak is already receding-- there is a response with a
significant hardening of the spectrum. \\
The fact that the spectral evolution at higher energies develops during the
decay phase of the flare, produces a nice counter-clockwise loop
in the $\alpha$ vs. Flux diagram, i.e. \textit{opposite} way with respect
to all the other known cases for HBLs (e.g. Sembay et al. 1993; Kohmura et
al. 1994; Takahashi et al.  1996).

\section*{Physical Interpretation}

Let us now focus on the possible interpretation of the two main results of
this work: the \textit{hard lag} and the \textit{evolution of the
synchrotron peak}. 
 
The occurrence of the flare peak at different times for different energies is
most likely related to the particle acceleration/heating process. 

We therefore \textit{introduced an acceleration term in the time
dependent particle kinetic equation} within the model proposed by
Chiaberge \& Ghisellini (1999), which takes into account the cooling and
escape terms and the role of delays in the received photons due to the
travel time from different parts of the emitting volume.

The main constraints on the (parametric) form of the acceleration are:
[A] particles have to be progressively accelerated from lower
to higher energies within the flare rise timescale to produce the hard lag; 
[B] the emission in the LECS band from the highest energy
particles (those radiating initially in the MECS band) should not exceed
that from the lower energy ones, as after the peak no further increase of
the (LECS) flux is observed;
[C] the total decay timescale might be dominated by the
achromatic crossing time effects, although the initial phase might be
partly determined by the different cooling timescales.

It should be also noted that --within this scenario-- the symmetry
between the raise and decay of the softer energy light curve seems to
suggest that at the same very energies where most of the power is
released --possibly determined by the balance between the
acceleration and cooling rates-- the acceleration timescales are comparable
to the region light crossing time.

If the timescales associated with this process are intrinsically linked to
the typical size of the emitting region, we indeed expect the observed
light curve to be symmetric where the bulk of power is concentrated, and an
almost achromatic decay.  

Indeed, within a single emission region
scenario, we have been able to reproduce the sign and amount of lags,
postulating that particle acceleration follows a simple law, and
stops at the highest particle energies.
The same model can account for the spectral evolution (shift of the
synchrotron peak) during the flare.

\section*{Conclusions}

These results provide us with several \textit{temporal} and \textit{spectral
constraints} on any model. 
In particular, they could possibly be the \textit{first direct signature of
the ongoing acceleration process}, progressively ``pumping'' electrons from
lower to higher energies.  
The measure of the delay provides a tight constraint on the timescale of the
acceleration mechanisms.  


A last crucial point is that our results support the 
possibility of the presence and role of quasi--stationary emission.
The short-timescale, large-amplitude variability events could be 
attributed to the development of new individual flaring components
(possibly maintaining a quasi-rigid shape), giving rise to a spectrum
outshining a more slowly varying emission.  
The decomposition in these two components might allow to determine the
nature and modality of the dissipation in relativistic jets.


\end{document}